# Comments on the Hartree-Fock description of the Hooke's atom and suggestion for an accurate closed-form orbital

Sébastien RAGOT

Laboratoire Structure, Propriété et Modélisation des Solides (CNRS, Unité Mixte de Recherche 85-80). École Centrale Paris, Grande Voie des Vignes, 92295 CHATENAY-MALABRY, FRANCE

E-mail: at_home@club-internet.fr

## Abstract

The ground-state Hartree-Fock (HF) wavefunction of the Hooke's atom is not known in closed form, contrary to the exact solution. The single HF orbital involved has thus far been studied using expansion techniques only, leading to slightly disparate energies. Therefore, the present letter aims at proposing alternative definitions of the HF wavefunction. First, the HF limit is ascertained using a simple expansion, which makes it possible to formulate explicit expressions of HF properties. The resulting energy, 2.038 438 871 8 $E_h$, is found stable at the tenth digit. Second and more instructive, an analysis of the Hartree equation makes it possible to infer a remarkably simple and accurate HF orbital, i.e. $\varphi_{HF}(r) = \mathcal{N}_t \, e^{-\alpha r^2} \sqrt{r^2 + \beta^2}$, leading to an energy exceeding by 5.76 $10^{-7}$ $E_h$ only the above HF limit. This orbital makes it possible to obtain (near) Hartree-Fock properties in closed-form, which in turn enables handy comparisons with exact quantities.

The Hookean model of the atom consists of two electrons harmonically trapped about a nucleus, which repel one another via a Coulomb term. The corresponding Hamiltonian is, in atomic units:

$$H = -\tfrac{1}{2}\nabla_1^2 - \tfrac{1}{2}\nabla_2^2 + \tfrac{1}{2}k(r_1^2 + r_2^2) + 1/r_{12}, \tag{1}$$

This model atom was introduced years ago by Kestner and Sinanoglu [1], willing to consider a tractable model of correlated electrons for which an exact solution can be formulated, at least in a power series. A ground-state wavefunction they formulated is

$$\psi = \mathcal{N} e^{-\frac{k^{1/2}}{2}(r_1^2 + r_2^2)} f(r_{12}), \tag{2}$$

wherein $f$ was shown to be almost linear in $r_{12}$. It has later been realized that the function $f$ is exactly $f(r_{12}) = 1 + r_{12}/2$ in the particular case of $k = 1/4$, whereby the exact wavefunction is obtained in closed-form [2]. The subsequent model ($k = 1/4$) has therefore logically attracted attention, enabling a number of properties of confined electrons to be derived in closed-form [3,4,5,6,7,8,9].

In contrast, the Hartree-Fock (HF) spatial wavefunction of the lowest singlet state, i.e. $\psi_{HF}(\mathbf{r}_1, \mathbf{r}_2) = \varphi_{HF}(r_1)\varphi_{HF}(r_2)$, is not known in closed-form, which impairs studies of correlation properties. Numerical approximations have nevertheless been investigated. In this respect, the single orbital $\varphi_{HF}(r)$ involved obeys the Hartree equation $h\varphi_{HF}(r) = \varepsilon_{HF}\varphi_{HF}(r)$, wherein $h$ reduces in this case to [4]:

$$\begin{aligned} h &= -\tfrac{1}{2}\nabla_r^2 + \frac{r^2}{8} + \int_0^\infty \frac{1}{r_{12}} \varphi_{HF}^2(r_2) d\mathbf{r}_2 \\ &= -\frac{1}{2}\frac{d^2}{dr^2} - \frac{1}{r}\frac{d}{dr} + \frac{r^2}{8} + \frac{1}{r}\int_0^r 4\pi r_2^2 \varphi_{HF}^2(r_2) dr_2 \\ &\quad + \int_r^\infty 4\pi r_2 \varphi_{HF}^2(r_2) dr_2 \end{aligned} \tag{3}$$

and the HF energy is given by:

$$E_{HF} = 2\int \varphi_{HF}(r_1)\left[-\frac{1}{2}\frac{d^2}{dr_1^2} - \frac{1}{r_1}\frac{d}{dr_1} + \frac{r_1^2}{8}\right]\varphi_{HF}(r_1)d\mathbf{r}_1 + \int_0^r \varphi_{HF}^2(r_1)\frac{1}{r_{12}}\varphi_{HF}^2(r_2)d\mathbf{r}_1 d\mathbf{r}_2. \qquad (4)$$

The Hartree equation, eq. (3), does *a priori* not allow for a simple solution. As an alternative, an expansion for the HF orbital allows for a variational calculation thereof. For instance, using a basis of harmonic oscillator eigenfunctions [4], O'Neill and Gill arrived at a limiting energy of 2.038 438 87 $E_h$.

Said value is singularly lower than previously reported HF energies, that is, 2.039 325 $E_h$ [10] or 2.038 51 $E_h$ [11], which have been obtained using *ab initio* programs suitably adapted to the harmonic potential. Using a similar approach, Katriel and co-workers have, however, obtained a value of 2.038 438 9 $E_h$ [12], in much closer agreement with ref. [4]. A consensus seems accordingly to emerge as to the value of 2.038 438 9 $E_h$.

However, such studies are all based on orbital expansions, which are not the most suited for a comparison with exact properties. Therefore, the purpose of this letter is to find an accurate approximation to the HF orbital in closed-form. To this aim, alternative definitions of the HF orbital are proposed. First, a simple expansion scheme is relied upon to ascertain the HF limiting energy and which makes it possible to formulate explicit expressions of HF properties. Then, a closed-form expression of the HF orbital is designed and compared to the most accurate orbital available.

In the following, HF energies are calculated by variationally minimizing analytic expression of the HF energy, as obtained from eq. (4). Numerical computations of energies have been carried out with *Mathematica* [13], by gradually increasing the number of digits of precision used in internal computations, so as to obtain final results stable to ten digits [14].

First, the HF orbital is expanded as

$$\varphi_{HF,n}(r) = \mathcal{N}_n e^{-\alpha r^2} p_n(r^2), \tag{5}$$

where $p_n(r^2)$ is a nth-order polynomial, intended to account for electron interaction. This orbital can be rewritten as

$$\varphi_{HF,n}(r) = \mathcal{N}_n \sum_{i=0}^{n} c_i (-1)^i \partial_\alpha^i e^{-\alpha r^2}, \tag{6}$$

where $\partial_\alpha^i \equiv \frac{\partial^i}{\partial \alpha^i}$ and $c_i$'s denote a set of n – 1 independent variational parameters, and

$$\mathcal{N}_n^{-2} = \sum_{i,j=1}^{n} c_i c_j (-1)^{i+j} \partial_\alpha^i \partial_\beta^j \left(\frac{\pi}{\alpha+\beta}\right)^{3/2}.$$

The HF energy as defined in eq. (4) becomes accordingly

$$E_{HF,n} = 2\mathcal{N}_n \sum_{i,j=0}^{n} c_i c_j (-1)^{i+j} E_{i,j}^{(1)}(\alpha) \tag{7}$$

$$+ \mathcal{N}_n^2 \sum_{i,j,k,l=0}^{n} c_i c_j c_k c_l (-1)^{i+j+k+l} E_{i,j,k,l}^{(2)}(\alpha),$$

where

$$E_{i,j}^{(1)}(\alpha) = \left[\partial_\alpha^i \partial_\beta^j \left(\frac{3\pi^{3/2}(1+16\alpha\beta)}{8(\alpha+\beta)^{5/2}}\right)\right]_{\beta=\alpha}, \text{ and} \tag{8}$$

$$E_{i,j,k,l}^{(2)}(\alpha) = \left[\partial_\alpha^i \partial_\beta^j \partial_\delta^i \partial_\gamma^j \left(\frac{2\pi^{5/2}}{(\alpha+\beta)(\delta+\gamma)(\alpha+\beta+\delta+\gamma)^{1/2}}\right)\right]_{\gamma=\beta=\delta=\alpha}.$$

A closed-form expression for the energy can thus be easily calculated thanks to eqs. (7) and (8), and then numerically brought to a minimum by optimizing parameters $c_n$. The exponent parameter $\alpha$ is preferably set to ¼ (as in the case of uncoupled oscillators), which makes high-order expansions easier to handle.

As to numerical results, the energies optimized for $n = 0$ to 6, eqs. (5)- (8), nicely match the results of O'Neill and Gill, table I of ref. [4], at least to the number of digits reported by the authors; detailed results are therefore not reported. A close agreement is, however, not surprising as the expansion used here merely amounts to using Hermite polynomials, as in ref. [4]. In particular, using a six-order polynomial results in 2.038 438 873 3 $E_h$, consistently with the value of correlation energy of $E_c = E_{exact} - E_h = 38.438\,873\,3$ m$E_h$ ($E_{exact} = 2\ E_h$) reported in ref. [4]. However, the HF limit that emerges from the present calculations is 2.038 438 871 8 $E_h$, to eleven digits of accuracy, using either $n = 9$, 10 or 11 in eqs. (5)- (8) [$E_{HF,11} = 2.038\,438\,871\,76\ E_h$ and $E_{HF,11} - E_{HF,10} < 10^{-11}$]. Accordingly, the correlation energy would be somewhat lower, i.e. 38.438 871 8 m$E_h$, than previously reported.

Hence, the HF orbital as defined in eq. (6) allows very accurate HF energies to be computed in a simple manner. Advantageously, since a single Gaussian is involved, eq. (6), any HF property can be expressed explicitly, including momentum-space or Wigner distributions.

However, since an expansion of the HF orbital is not very instructive *per se*, it is desirable to investigate a closed-form approximation thereof. To this aim, one may analyze the one-electron operator $h$ in the Hartree equation, eq. (3), and search approximations at both small and large values of $r$.

At small $r$, a limited development of the integrals in eq. (3), that is, of the Hartree potential, results in

$$h \approx -\tfrac{1}{2}\nabla_r^2 + \frac{k'r^2}{2} + \vartheta(r^4), \tag{9}$$

wherein $k' = k - \tfrac{4}{3}\pi[\varphi_{HF}(0)]^2$. Thus, the effective potential reduces in this case to that of an uncoupled harmonic oscillator, with a modified constant. As the potential in eq. (9) does not depend on $r^3$, the HF orbital must depend quadratically on $r$, correct to third order. Basically, using $k = 0.25$ and $\varphi_{HF,9}$ as obtained above, the numerical value that raises is about $k' = 0.06$. This shows that the mean-field has a substantial effect at small $r$, making an electron loosely bound at the origin compared with uncoupled oscillators.

In contrast, at large $r$, the integral terms in eq. (3) can be recast as

$$\frac{1}{r}\int_0^r 4\pi r_2^2 \varphi_{HF}^2(r_2)dr_2 + \int_r^\infty 4\pi r_2 \varphi_{HF}^2(r_2)dr_2$$

$$= \frac{1}{r} + \int_r^\infty 4\pi r_2^2 \varphi_{HF}^2(r_2)\left[\frac{1}{r_2} - \frac{1}{r}\right]dr_2$$

$$\approx \frac{1}{r}.$$

Thus, the one-electron operator $h$ becomes approximately

$$h \approx -\tfrac{1}{2}\nabla_r^2 + \frac{kr^2}{2} + \frac{1}{r}, \qquad (10)$$

which is in fact a rescaled version of the intracular component of the two-particle Hamiltonian, namely $H_{r_{12}} = -\nabla_{r_{12}}^2 + \tfrac{1}{4}kr_{12}^2 + 1/r_{12}$. This becomes apparent when substituting $k \to 4k$ and $r \to r_{12}/2$ in eq. (10) above, and comparing the result with e.g. eq. (6) of ref. [1]. The solution discussed in ref. [1] suggests that $\varphi_{HF}(r)$ behaves at large $r$ approximately as

$$\varphi_{HF}(r) \propto e^{-\frac{k^{1/2}}{2}r^2}(1+r+...) \approx e^{-\frac{k^{1/2}}{2}r^2} r.$$

Therefore, a good candidate for a trial HF orbital, having the correct behaviors at both small and large values of $r$, is

$$\varphi_{HF}(r) = \mathcal{N}_{HF}\, e^{-\alpha r^2}\sqrt{r^2+\beta^2}, \qquad (11)$$

where

$$\mathcal{N}_{HF} = \frac{\left(2(2/\pi)^{3/4}\alpha^{5/4}\right)}{\sqrt{3+4\alpha\beta^2}}.$$

The corresponding charge density is

$$\rho_{HF}(r) = 2\mathcal{N}_{HF}^2\, e^{-2\alpha r^2}(r^2+\beta^2), \qquad (12)$$

and the Hartree potential is

$$v_h(r) = \mathcal{N}_{HF}^2 \frac{\pi}{16\alpha^{5/2}}\left(-4\sqrt{\alpha}\, e^{-2\alpha r^2} + \sqrt{2\pi}(3+4\alpha\beta^2)\frac{\text{erf}(\sqrt{2\alpha}\, r)}{r}\right). \quad (13)$$

The Hartree potential is maximum at the origin and slowly decreases towards large $r$ values. function. Accordingly, the mean-field approximation results in substantially redistributing electrons farther to the nucleus. Similarly, the pair density and kinetic energy densities can be obtained in closed-form.

Inserting now the trial orbital in eq. (4) yields an expression for the HF energy:

$$E_{HF} = \quad (14)$$
$$\frac{1}{2(3+4\alpha\beta^2)^2}\Big\{\sqrt{\alpha/\pi}\big(27+16\alpha\beta^2(5+4\alpha\beta^2)\big)+$$
$$\frac{1}{8\alpha}(3+4\alpha\beta^2)\big[15+4\alpha(3\beta^2+4\alpha(7-4\alpha\beta^2))\big]+$$
$$256\sqrt{2\pi\alpha}(\alpha\beta)^3 e^{2\alpha\beta^2}\text{erfc}(\sqrt{2\alpha}\,\beta)\Big\},$$

wherein erfc($x$) is 1 - erf($x$). A relevant value of $\beta$ can further be obtained by enforcing the virial theorem, i.e. $2T = \langle \psi | \sum_{j=1,2} \mathbf{r}_j . \nabla_j v(\mathbf{r}_1,\mathbf{r}_2) | \psi \rangle$, which in the present case leads to

$$2T_{HF} = \int \rho_{HF}(\mathbf{r})\mathbf{r}.\nabla(v_{ext}(\mathbf{r}) + v_h(\mathbf{r}))d\mathbf{r} \quad (15)$$
$$= 2U_{HF,ext} - U_{HF,h},$$

where $U_{HF,ext}$ and $U_{HF,h}$ are the external and Hartree energies, respectively. Using $\alpha = k^{1/2}/2 = \frac{1}{4}$ (as for uncoupled oscillators), the energy components at stake evaluates to

$$U_{HF,h} = \frac{4(\beta^2+5)\beta^2+27}{4\pi^{1/2}(3+\beta^2)^2},$$

$$U_{HF,ext} = \frac{3}{4} + \frac{3}{2(3+\beta^2)}, \text{ and}$$

$$T_{HF} = \frac{2(2\pi)^{1/2}\beta^3 e^{\beta^2/2}\text{erfc}(2^{-1/2}\beta) - \beta^2 + 7}{4(3+\beta^2)}.$$

Numerically solving eq. (15) leads to $\beta = 2.771\,219\,931$, whereby the HF trial orbital is now fully parameterized. As such, it yields an energy which is 4.5 $10^{-6}$ $E_h$ above the HF limit.

Keeping in mind this result is obtained from a parameterized orbital (not yet optimized), the inferred HF orbital seems to be remarkably accurate.

Interestingly, the exact wavefunction can be simply rewritten in terms of HF orbitals parameterized as above, that is $\psi = \varphi_{HF}(r_1)\varphi_{HF}(r_2)g(r_{12})$, wherein the correlation factor $g$ takes a very simple form, i.e.

$$g(r_{12}) = \frac{\mathcal{N}}{\mathcal{N}_{HF}^2} \frac{(1 + r_{12}/2)}{\sqrt{r_1^2 + \beta^2}\sqrt{r_2^2 + \beta^2}}. \tag{16}$$

The simplicity of such a correlation factor may advantageously find applications in calculations based on empirically correlated atomic wavefunctions [15,16].

Finally, relaxing both parameters of the trial orbital $\varphi_{HF}(r)$ in eq. (14) leads to an energy of 2.038 439 449 1 $E_h$, that is, in excess of about 5.77 $10^{-7}$ $E_h$ only with respect to the most accurate HF energy available. The optimal parameters are $\alpha = 0.251\ 117\ 376$ (closed to the "uncoupled" value, $\alpha = 1/4$) and $\beta = 2.711\ 087\ 898$.

The accuracy of this optimized orbital $\varphi_{HF}(r)$ can be assessed in various manners. First, it is visually not distinguishable from the most accurate orbitals tested ($\varphi_{HF,9} - \varphi_{HF,11}$), even in radial form. The largest difference with the limit HF orbital occurs at the origin and is of about 1.7 $10^{-4}$, the largest difference between the corresponding charge density being of the same magnitude. Second, though the optimized orbital $\varphi_{HF}(r)$ is not a true eigenvector of the operator $h$, eq. (3), the effective eigenvalue $\varepsilon_{HF}(r) = h\varphi_{HF}(r)/\varphi_{HF}(r)$ that raises is essentially flat. For instance, the standard deviation of $\varepsilon_{HF}(r)$ in a statistical sense, i.e. taken as $\sigma_{\varepsilon_{HF}} = \sqrt{\langle \varphi_{HF} |(\varepsilon_{HF}(r)^2 - \bar{\varepsilon}_{HF}^2)| \varphi_{HF}\rangle}$, is numerically found to be of about 1.1 $10^{-3}$, while the value of $\bar{\varepsilon}_{HF} = \langle \varphi_{HF} |h| \varphi_{HF}\rangle$ is 1.276 681 $E_h$.

To summarize this short study of the HF description of the Hooke's atom, variational calculations based on alternative definitions of the HF orbital, eqs. (4) and (7), merely confirm the lowest HF limit reported to date [4], though accuracy is somewhat improved. All the more, a brief analysis of the Hartree equation suggests a remarkably accurate HF orbital, $\varphi_{HF}(r) = \mathcal{N}_{HF}\, e^{-\alpha r^2} \sqrt{r^2 + \beta^2}$, enabling relevant HF properties to be evaluated in closed-form. Said orbital can be considered as the exact HF orbital, as long as the precision sought in energy is not better than $5.77\ 10^{-7}\ E_h$, which is more than required for most applications.